# Brief Report

neuroscience

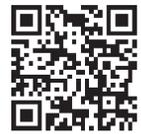

# mGluR5 knockout mice exhibit normal conditioned place-preference to cocaine

**Melissa A. Fowler[2], Andrew L. Varnell[1] & Donald C. Cooper[1*]**

**Metabotropic glutamate receptor 5 (mGluR5) null mutant (-/-) mice have been reported to totally lack the reinforcing or locomotor stimulating effects of cocaine. We tested mGluR5 -/- and +/+ mice for their locomotor and conditioned place- preference response to cocaine. Unlike the previous finding, here we show that compared to mGluR5 +/+ mice, -/- mice exhibit no difference in the locomotor response to low to moderate doses of cocaine (10 or 20 mg/kg). A high dose of cocaine (40 mg/kg) resulted in a blunted rather than absent locomotor response. We tested mGluR5 -/- and +/+ mice for conditioned place-preference to cocaine and found no group differences at a conditioning dose of 10 mg/kg, suggesting normal conditioned rewarding properties of cocaine. These results differ substantially from Chiamulera et al. (2001) and replicates Olsen et al., (2010), who found normal cocaine place-preference in mGluR5 -/- mice at 5 mg/kg. Our results indicate mGluR5 receptors exert a modulatory rather than necessary role in cocaine-induced locomotor stimulation and exert no effect on the conditioned rewarding effects of cocaine.**

Metabotropic glutamate receptor 5 (mGluR5) is Gq-coupled and, when activated by glutamate, initiates a signaling cascade that increases phospholipase C (PLC) and elevates intracellular $Ca^{2+}$ from the endoplasmic reticulum which then culminates in the activation of nonselective cation TRPC4/5 currents[1,2]. Expression of mGluR5 in the brain is widespread, with some of the highest expression in the hippocampus, prefrontal cortex and striatum. These areas are important for mediating the behavioral responses to psychostimulant drugs.

Behavioral pharmacology studies have shown that administration of mGluR5 antagonists reduce cocaine self-administration and reinstatement[3]. The most compelling support for a necessary role for mGluR5 in cocaine mediated behaviors came from Chiamulera et al. (2001) who used mGluR5 null mutant (-/-) mice to test the locomotor activating and rewarding properties of cocaine[4]. These authors reported a complete absence of locomotor response across a wide dose response range (10-40 mg/kg, i.p.) in the mGluR5 -/- mice[3]. Furthermore, they reported a total lack of cocaine self-administration (SA) across all doses (0.08 -3.2 mg/kg i.v.) in the -/- mice despite normal responding for sucrose reward. It would appear from this study that cocaine is not acutely rewarding in mice lacking the mGluR5 receptor, however, it may be that the conditioned rewarding effects of cocaine are intact.

A protocol that is widely used to test the conditioned rewarding effects of cocaine is the conditioned place-preference (CPP) task[1]. In this task mice are repeatedly injected with either saline or cocaine and confined to an associated context. After a few pairings the mice are allowed to choose which context they prefer under drug-free conditions. Using this protocol mice prefer the cocaine-paired context across a range of doses usually between 5 and 10 mg/kg. To examine whether mGluR5 is necessary for this type of cocaine mediated reward we tested CPP in +/+ and -/- mice.

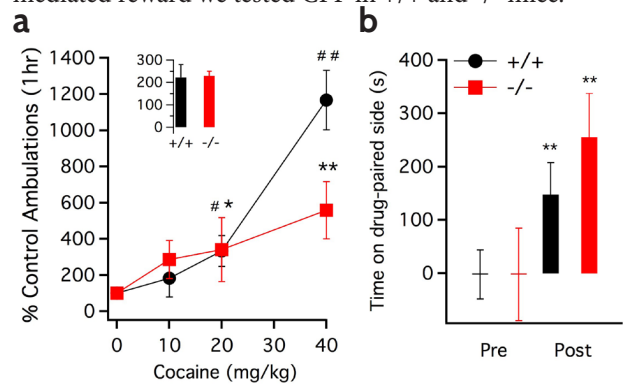

Fig. 1: **(a)** mGluR5 -/- mice show locomotor activation to cocaine at 20 mg/kg (*$p < 0.05$) and 40 mg/kg (**$p < 0.01$) doses. The increased ambulatory locomotor response to i.p. injections was similar in +/+ mice at 10 and 20 mg/kg (#$p < 0.05$) doses. At 40 mg/kg cocaine, mGluR5 +/+ (n=8) locomoted significantly more than -/- (n=5) mice ($p < 0.05$). (Inset) Locomtor response to saline injection was not different between +/+ and -/- mice. **(b)** Loss of mGluR5 does not reduce cocaine conditioned place preference at a 10 mg/kg conditioning dose. Cocaine caused significant conditioned place-preference at 10 mg/kg of cocaine in mGluR5 +/+ (**$p < 0.01$; n=13) and -/- (**$p < 0.01$; n=14) mice.

## RESULTS

*Locomotor response to cocaine*
While previous studies have shown that mGluR5 -/- mice have no ambulatory locomotor response to cocaine at doses of 10, 20 and 40 mg/kg compared to mGluR5 +/+ mice, our results failed to replicate this finding at low to moderate doses of 10 and 20 mg/kg. However, there was a significant decrease in the ambulatory movement of mGluR5 -/- mice at 40 mg/kg cocaine compared to mGluR5 +/+.

1. Center for Neuroscience, Department of Psychology and Neuroscience, Institute for Behavioral Genetics, University of Colorado at Boulder, Boulder Colorado 80303, USA
2. Johns Hopkins School of Medicine, Department of Department of Biological Chemistry
*Correspondence should be sent to D.Cooper@Colorado.edu

*Cocaine conditioned place-preference*

Prior cocaine SA studies have shown that mGluR5 -/- mice fail to SA cocaine at any dose compared to mGluR5 +/+ mice[2]. SA is a behavioral paradigm used to directly measure the rewarding properties of cocaine. Alternative methods, such as CPP, have been developed to measure the conditioned reinforcing properties of cocaine. The CPP procedure involves repeated noncontingent injections of cocaine to mice paired with a specific context and explicit unpairing in a different context. Then, in the absence of the drug, the mice are allowed to move freely between the two contexts. The amount of time spent in the cocaine-paired and unpaired contexts is the indirect measure of the rewarding property of cocaine. CPP is complementary to SA because it assesses cocaine reward memory in the absence of the locomotor activating properties of cocaine. Before conditioning, neither group showed preference to the cocaine- paired side. After conditioning to a 10 mg/kg dose of cocaine, both mGluR5 -/- and +/+ mice showed significant preference for the cocaine-paired side and there were no group differences (Fig. 1).

## DISCUSSION

Chiamulera et al. (2001) tested cocaine SA and found that the mGluR5 -/- mice fail to SA or ambulate to cocaine at any dose[3]. In contrast, our results indicate that mGluR5 -/- mice exhibit normal locomotor responses to low to moderate doses (10 and 20 mg/kg, i.p.) of cocaine. However, we did observe a reduced locomotor response to a high dose (40 mg/kg, i.p.) of cocaine compared to +/+ mice. To measure the ability of mGluR5 -/- mice to form rewarding drug-contextual associations to cocaine we tested cocaine CPP. Once the drug context association is formed through repeated pairings of the drug with a particular context, the animal is allowed to choose between the drug-paired context and an unpaired context. If the drug is considered rewarding, the animals will spend the majority of the time in the drug-paired context. Our results using a 10 mg/ kg (i.p.) dose of cocaine replicate Olsen et al., (2010)[4] who found normal CPP in mGluR5 -/- mice using a 5 mg/kg dose. Together these findings indicate that compared to mGluR5 +/+ mice, the -/- mice show normal CPP to cocaine. This is interesting given the Chiamulera et al., (2001)[3] report indicating that mGluR5 -/- mice do not self-administer or locomote to cocaine across a wide dose range from 0.08-3.2 mg/kg (i.v) for self-administration and 10-40 mg/kg (i.p) for locomotor activity.

It is possible that mGluR5 receptors are not involved in the drug-paired conditioned rewarding effects of cocaine while the direct rewarding property of cocaine relies on mGluR5. Although McGeehan et al., (2003) found that pharmacological antagonism of mGluR5 receptors with MPEP >5 mg/kg reduced cocaine CPP[5]. Regional differences in the expression of mGluR5 may account for the difference between cocaine CPP and SA. Studies in our laboratory are underway reexamining cocaine SA in mGluR5 -/- mice and updates to the project will be posted on www.neuro-cloud.net[1].

## METHODS

*Locomotor response to cocaine*

Adult male mGluR5 +/+ and -/- mice were weighed and marked one day prior to the start of the behavioral testing. Mice were habituated to the testing room for one hour prior to the start of testing on each day of testing. The locomotor chambers consisted of a cage identical to the home cage placed into a box equipped with five photobeams for recording movement. Ambulatory movement was defined as two consecutive beam breaks. On days one and two, after one hour of habituation to the chamber, the animals were injected with saline and placed back into the chambers where movement was recorded for an additional hour. The animals were then given ascending doses of cocaine of 10, 20 and 40 mg/kg. A saline injection was interleaved between the cocaine injections to prevent conditioned locomotion.

*Cocaine conditioned place-preference*

Adult male mGluR5 +/+ and -/- mice were weighed and marked one day prior to the start of the behavioral procedure. The CPP chamber consisted of two large compartments that differed in both visual and tactile cues, separated by a smaller center compartment (Med Associates, Vermont, Georgia). The two large compartments were equipped with six photo beams for monitoring movement, and the center compartment with two photobeams. On training days one through four, partitions were used to block off the two large compartments and each animal was given an i.p. Injection of either cocaine or saline (10 mg/kg alternate over four days) and placed into its respective compartment for 30 min. On the fifth day, the partitions were lifted and the animals were placed into the center compartment. The animals were allowed to move freely between the three compartments for 20 min. and the time spent in each compartment was recorded. The preference for the cocaine-paired side was calculated as the time spent in the cocaine-paired side minus the time spent in the saline-paired side. For expanded methods and recordings please see http:// www.neuro-cloud.net/nature-precedings/fowler/ .

**PROGRESS AND COLLABORATIONS**

To see up to date progress on this project or if you are interested in contributing to this project visit: http://www.neuro-cloud.net/nature-precedings/fowler/

**AUTHOR CONTRIBUTIONS**

M.A.F. and D.C.C. Designed the Experiment, M.A.F. recorded and analyzed the data, A.L.V.. M.A.F. and D.C.C. prepared the manuscript.

**ACKNOWLEDGEMENTS**

We thank K. Huber (University of Texas Southwestern Medical Center at Dallas) for the mGluR5 wild-type and knockout mice. This work was supported by National Institute on Drug Abuse grant R01-DA24040 (to D.C.C.), NIDA K award K-01DA017750 (to D.C.C.), a NARSAD Young Investigator award (to D.C.C.), National Institute on Drug Abuse institutional training grant T32-DA7290 (to M.A.F.). We would like to acknowledge funding in part from the University of Texas Southwestern Medical Center and the University of Colorado, Boulder. We would like to acknowledge financial support and helpful editing discussion from Leah Leverich Ph.D.

Submitted online at http://www.precedings.nature.com